\documentclass[fleqn,10pt]{wlscirep}
\begin{document}
\title{The classical correlation limits the ability of the measurement-induced average coherence}
 \author{Jun Zhang}
 \affil{College of Mathematics, Institute of Mathematics, Taiyuan University of Technology, Taiyuan 030024, China}
 \author{Si-ren Yang}
 \author{Yang Zhang}
 \author[2,*]{Chang-shui Yu}
\affil{School of Physics and Optoelectronic Technology, Dalian University of Technology, Dalian 116024, China }
\affil[*]{quaninformation@sina.com;ycs@dlut.edu.cn}
\begin{abstract}
Coherence is the most fundamental quantum feature in quantum mechanics. For a bipartite quantum state, if a measurement is performed on one party, the other party, based on the measurement outcomes, will collapse to a corresponding state with some probability and hence gain the average coherence. It is shown that the average coherence is not less than the coherence of its reduced density matrix. In particular, it is very surprising that the extra average coherence (and the maximal extra average coherence with all the possible measurements taken into account) is upper bounded by the classical correlation of the bipartite state instead of the quantum correlation. We also find the sufficient and necessary condition for the null maximal extra average coherence. Some examples demonstrate the relation and, moreover, show that quantum correlation is neither sufficient nor necessary for the  nonzero extra average coherence within a given measurement. In addition, the similar conclusions are drawn for both the basis-dependent and the basis-free coherence measure.
\end{abstract}
\flushbottom
\maketitle
\thispagestyle{empty}
\section*{Introduction}

Quantum coherence originating from the quantum superposition principle is the most fundamental quantum feature of quantum mechanics. It plays an important role in various fields such as  the thermodynamics \cite{Aberg113,Narasimhachar6,Cwiklinski115,Lostaglio6, Scully108,Scully299} , the transport theory \cite{levi,reb,witt,Wang53} , the living complexes \cite{Plenio10,Lloyd302,Huelga54} and so on. With the resource-theoretic understanding of quantum feature in quantum information, the quantification of coherence has attracted increasing interest in recent years \cite{Baumgratz113,Girolami113,Pires91,Shao91,Rana93,Zhang93} and has also led to the operational resource theory of the coherence \cite{Winter116} .

The quantitative theory also makes it possible to understand one type of quantumness (for example, the coherence) by the other type of quantumness such as the entanglement and the quantum correlation, \textit{vice versa} \cite{Yu80,Hu,Chitambar116,tan,Yu13,Yao92,Xi5,Cheng92,Singh91, Singh93,Du91} . For example, for a bipartite pure state, the maximal extra average coherence that one party could gain was shown to be exactly characterized by the concurrence  assisted by the local operations and classical communication (LOCC) with the other party \cite{Yu80} .  Ref. \cite{Hu} showed that the maximal average coherence was bounded by some type of quantum correlation in some particular reference framework. In the asymptotic regime, Ref. \cite{Chitambar116} showed that the  rate of assisted coherence distillation
for pure states was equal to the coherence of assistance under the local quantum-incoherent operations and classical communication. Quite recently, a unified view of quantum correlation and quantum coherence has been given in Ref. \cite{tan} . In addition, if only the incoherent operations are allowed, the state with certain amount of coherence assisted by an incoherent state can be converted to an entangled state with the same amount of entanglement \cite{Streltsov115} or a quantum-correlated state with the same amount of quantum correlation \cite{Ma116} .

In this paper, \textit{instead of the quantum correlation}, we find, it is  \textit{the classical correlation} of a bipartite quantum state that limits the extra average coherence at one side induced by the unilateral measurement at the other side.  We also find the necessary and sufficient condition for the zero maximal average coherence that could be gained with all the possible measurements taken into account. Besides, we show, through some examples, that quantum correlation is neither sufficient nor necessary  for the extra average coherence subject to a given measurement. We have selected both the basis-dependent and the basis-free coherence measure to study this question and obtain the similar conclusions. In particular, one should note that all our results are valid for the positive-operator-valued measurement (POVM), even though we only consider the local projective measurement in the main text.

\section*{Results}

\subsection*{The upper bound on the extra measurement-induced average coherence}

\textbf{{Coherence measure}-} To begin with, let's first give a brief review of the measure of the quantum coherence \cite{Baumgratz113} . If a  quantum state $%
\hat{\delta}$  can be written as
\begin{eqnarray}
\hat{\delta}=\sum_{i}\delta_{i}\left\vert i\right\rangle\left\langle i\right\vert,
\end{eqnarray}
$\hat{\delta}$ is incoherent with respect to the basis $\{\left\vert i\right\rangle\}$. Let $\mathcal{I}$ denotes the set of incoherent states, then
the operator $\hat{K}_{n}$  is the incoherent operation if it satisfies  $\hat{K}_{n}\mathcal{I}\hat{K}_{n}^{\dagger}\subset
\mathcal{I}$. Thus a
good coherence measure  $C(\rho)$ of a $d$-dimensional state $\rho$ should be:

(p1) Nonnegative-i.e., $C(\rho)\geq 0$ and $C(\rho)= 0$ if and only if the
quantum state $\rho$ is incoherent.

(p2) Monotonic-i.e., $C(\rho)\geq C(\Lambda(\rho))$ for any incoherent
operation $\Lambda(\rho)=\sum_{n}\hat{K}_{n}\rho\hat{K}_{n}^{\dagger}$;
and strongly monotonic if $C(\rho)\geq \sum_{n}p_{n}C(\rho_{n})$ with $%
p_{n}\rho_{n}=\hat{K}_{n}\rho\hat{K}_{n}^{\dagger}$.

(p3) Convex-i.e., $C(\sum_{i}p_{i}\rho_{i})\leq \sum_{i}p_{i}C(\rho_{i})$.

 Even though there are many good coherence measures such as the coherence measures based on $l_{1}$-norm, trace norm, fidelity,
 the relative entropy and so on \cite{Baumgratz113,Girolami113,Pires91,Shao91,Rana93,Zhang93} , in this paper we will only employ the relative entropy to quantify the quantum coherence, i.e.,
\begin{eqnarray}
{\mathcal{C}}(\rho)&=&\min_{\delta\in \mathcal{I}} S(\rho\|\delta)=S(\rho^\star)-S(\rho).
\end{eqnarray}
where $S(\rho\|\sigma)=\mathrm{Tr}\rho\log\rho-\mathrm{Tr}\rho\log\sigma$ is the relative entropy, $S(\rho)=-\mathrm{Tr}\rho\log\rho$ is the von Neumann entropy and  $\rho^\star$ is the diagonal matrix by deleting all the off-diagonal entries of any $\rho$ (we will use this notation throughout the paper).
For simplicity, we will restrict ourselves in the computational basis throughout the paper. In contrast, the basis-free coherence (or the total coherence) \cite{Yuquantum}  is quantified by
\begin{eqnarray}
{\mathcal{C}}^{T}(\rho)=S(\rho\|\frac{\mathbb{I}_{d}}{d})=\log d-S(\rho).
\end{eqnarray}
Note that $\mathcal{C}^{T}(\rho)$ quantifies the maximal coherence of a state with all the bases taken into account.
\begin{figure}[t]
\centering
\includegraphics[width=7cm]{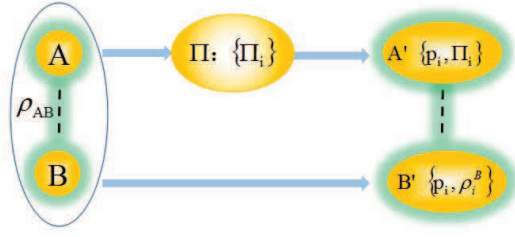}\newline
\caption{(Color online) Illustration of the two-player game on the measurement-induced average coherence.}
\label{fig1}
\end{figure}

\textbf{{The Classical correlation as the upper bound}-} Now let's turn to our game  sketched in FIG. 1. Suppose two players, Alice and Bob, share a two-particle quantum state $\rho_{AB}$ and Alice performs some projective measurement $\Pi:\{\Pi_i\}$ on her particle and sends her outcomes to Bob. Bob isn't allowed to do any operation. Based on Alice's outcomes, Bob will obtain the state $ \rho^B_i=(\Pi^A_i\otimes \mathbb{I}_B)\rho_{AB}(\Pi^A_i\otimes \mathbb{I}_B)/p_i$
with the probability $p_i=\mathrm{Tr}(\Pi^A_i\otimes \mathbb{I}_B)\rho_{AB}(\Pi^A_i\otimes \mathbb{I}_B)$.
Thus in the computational basis, the measurement-induced average coherence (MIAC: Bob's average coherence induced by Alice's measurement $\Pi$) is given by
\begin{equation}
\overline{\mathcal{C}}^{P}_{\Pi}(\rho _{B})=\sum_{i}p_{i}{\mathcal{C}}(\rho_i^B)=\sum_{i}p_{i}\min_{\delta_i \in \mathcal{%
I}}S(\rho ^{B}_{i}\Vert \delta ^{B}_i).\label{miac}
\end{equation}%
Similarly, the measurement-induced average total coherence (MIATC: Bob's average total coherence induced by Alice's measurement $\Pi$) is
\begin{eqnarray}
\overline{\mathcal{C}}_{\Pi}^{T}(\rho_{B})=\sum_{i}p_{i}{\mathcal{C}}^T(\rho_i^B)=\log d-\sum_{i}p_{i}S(\rho^{B}_{i}), \label{miatc}
\end{eqnarray}
with $d$ denoting the dimension of Bob's space. With Alice's measurement $\Pi$, the Bob's average coherence is usually different from the coherence of  $\rho_{B}=\mathrm{Tr}_A\rho_{AB}$.The extra MIAC $(\Delta\mathcal{C}_{\Pi}^{P})$ and the extra MIATC $(\Delta\mathcal{C}_{\Pi}^{T})$ can be defined as
\begin{eqnarray}
\Delta\mathcal{C}_{\Pi}^{P}&=&\overline{\mathcal{C}}_{\Pi}^{P}(\rho_{B})-{\mathcal{C}}(\rho_{B}),\\
\Delta\mathcal{C}_{\Pi}^{T}&=&\overline{\mathcal{C}}_{\Pi}^{T}(\rho_{B})-{\mathcal{C}^{T}}(\rho_{B}).
\end{eqnarray}
It is obvious that $\Delta\mathcal{C}_{\Pi}^{P/T}\geq 0$ which is impied by the convexity of the coherence $\mathcal{C}$, that is,  $\sum_ip_i\mathcal{C}(\rho_i^B)\geq\mathcal{C}(\rho_B)$ with $\rho_B=\sum_ip_i\rho_i^B$.

Thus our main results can be given by the following theorems.

\textbf{Theorem 1}: For a bipartite quantum state $\rho_{AB}$, the extra MIAC $\Delta\mathcal{C}_{\Pi}^{P}$ is not greater than the extra $\Delta\mathcal{C}_{\Pi}^{T}$, i.e.,
\begin{equation}
\Delta\mathcal{C}_{\Pi}^{P}\leq \Delta\mathcal{C}_{\Pi}^{T}.
\end{equation}

\emph{Proof.} Based on Eq. (\ref{miac}), we have
\begin{eqnarray}
\overline{\mathcal{C}}^{P}_{\Pi}(\rho _{B})&=&\sum_{i}p_{i}{\mathcal{C}}(\rho_i^B)=\sum_{i}p_{i}\min_{\delta_i^B \in
\mathcal{I}}S(\rho ^{B}_{i}\Vert \delta_i ^B)  \notag \\
&\leq &\sum_{i}p_{i}S(\rho ^{B}_{i}\Vert \rho _{B}^\star)  \notag \\
&=&S(\rho_B)-\sum_{i}p_{i}S(\rho ^{B}_{i})\notag\\
&&-S(\rho_B)-\mathrm{Tr}\sum_{i}p_{i}\rho ^{B}_{i}\log
\rho _{B}^\star  \notag \\
&=&S(\rho _{B})-\sum_{i}p_{i}S(\rho ^{B}_{i}) +\mathcal{C}(\rho_B), \label{MIATCDA}
\end{eqnarray}%
with $\rho_B=\mathrm{Tr}_B\rho_{AB}$. Substituting the definition of MIATC (Eq. (\ref{miatc})) into Eq.(\ref{MIATCDA}), we can obtain the
\begin{eqnarray}
 \Delta\mathcal{C}_{\Pi}^{P}&=&\overline{\mathcal{C}}^{P}_{\Pi}(\rho _{B})-\mathcal{C}(\rho_B)\notag\\
 &\leq& S(\rho _{B})+\overline{\mathcal{C}}_{\Pi}^{T}(\rho_{B})-\log d \notag\\
  &=& \overline{\mathcal{C}}_{\Pi}^{T}(\rho_{B})-[\log d-S(\rho_{B})] \notag\\
&=& \Delta\mathcal{C}_{\Pi}^{T}.
\end{eqnarray}
The inequality holds if all Bob's states $\rho_{B}$ and $\rho_{i}^{B}$ have the same diagonal entries. The proof is completed. \hfill $\blacksquare$

\textbf{Theorem 2}: For a bipartite quantum state $\rho_{AB}$, the extra MIAC $\Delta\mathcal{C}_{\Pi}^{P}$ is upper bounded by
the classical correlation of $\rho_{AB}$, that is,
\begin{equation}
\Delta\mathcal{C}_{\Pi}^{P}\leq\mathcal{J}(B|\{\Omega _{i}^A\}),\label{zong}
\end{equation}
where the classical correlation is defined by
\begin{equation}
\mathcal{J}(B|\{\Omega ^A _{i}\})=S(\rho _{B})-\min_{\{\Omega ^A _{i}\}}S(B|\{\Omega _{i}^{A}\}), \label{jingdian}
\end{equation}
with $S(B|\{\Omega  _{i}^{A}\})=\sum_{i}q_{i}S(\varrho _{i}^{B})$ and $\{q_{i},\varrho _{i}^{B}\}$ defined by
\begin{equation}
\varrho^B_i=(\Omega^A_i\otimes \mathbb{I}_B)\rho_{AB}(\Omega^A_i\otimes \mathbb{I}_B)/q_i, \label{state}
\end{equation}
and the corresponding probability
\begin{equation}
q_i=\mathrm{Tr}(\Omega^A_i\otimes \mathbb{I}_B)\rho_{AB}(\Omega^A_i\otimes \mathbb{I}_B).\label{proba}
\end{equation}
Eq. (\ref{zong}) saturates if $\rho _{i}^{B}$ induced by the measurement $\Pi$ achieves the classical correlation $\mathcal{J}(B|\{\Omega ^A _{i}\})$ and $(\rho _{i}^{B})^\star$'s  are the same for all $i$. An example is
 the pure state $|\varphi\rangle_{AB}=\sum_{j}\lambda_{j}U_{A}\otimes\mathbb{I}_B|j\rangle_{A}|\tilde{j}\rangle_{B}$ where $U_{A}$ is unitary, $\left\vert j\right\rangle$, $\left\vert \tilde{j}\right
\rangle$ are the local computational basis.

\textbf{Theorem 3}:  The extra MIATC $\Delta\mathcal{C}_{\Pi}^{T}$ for a bipartite quantum state $\rho_{AB}$ is upper bounded by the classical correlation $\mathcal{J}(B|\{\Omega _{i}^A\})$ of $\rho_{AB}$, i.e.,
\begin{equation}
\Delta\mathcal{C}_{\Pi}^{T}\leq\mathcal{J}(B|\{\Omega _{i}^A\}).\label{miatct}
\end{equation}
 The equality holds for the pure $\rho_{AB}$.

\emph{Proof.} From the classical correlation, we have
\begin{eqnarray}
\mathcal{J}(B|\{\Omega ^A _{i}\})&=&S(\rho _{B})-\min_{\{\Omega _{i}\}}\sum_{i}q_{i}S(\varrho _{i}^{B})   \notag \\
&\geq& S(\rho _{B})-\sum_{i}p_{i}S(\rho _{i}^{B}).
\label{ljy}
\end{eqnarray}
Substituting Eq. (\ref{miatc}) into Eq. (\ref{ljy}), one can arrive at
\begin{eqnarray}
 \mathcal{J}(B|\{\Omega ^A _{i}\})&\geq&-\log d+S(\rho_{B})+\overline{\mathcal{C}}_{\Pi}^{T}(\rho_{B}) \notag \\
&= &\overline{\mathcal{C}}_{\Pi}^{T}(\rho_{B})-[\log d-S(\rho_{B})] \notag\\
&=&\Delta\mathcal{C}_{\Pi}^{T}.
\end{eqnarray}
Since both  $\mathcal{J}(B|\{\Omega ^A _{i}\})=S(\rho _{B})$ and $S(\rho_i^B)=0$ hold for pure $\rho_{AB}$ \cite{Henderson34} ,  the inequality (\ref{ljy}) saturates for the pure quantum state $\rho_{AB}$.
The proof is finished. \hfill $\blacksquare$

All the above three theorems hold for any projective measurement, so if we specify the particular measurement such that the maximal extra MIAC or MIATC can be achieved, the three theorems are also valid, which can be given in a rigorous way as:

\textbf{Corollary 1}. For a bipartite state $\rho_{AB}$ with the reduced density matrix $\rho_B$, the maximal extra MIAC and the maximal extra MIATC satisfy
\begin{equation}
\Delta\mathcal{C}^{P}_{\max}\leq\Delta\mathcal{C}^{T}_{\max},
\end{equation}
and
\begin{eqnarray}
\Delta\mathcal{C}^{P}_{\max}&=&\overline{\mathcal{C}}^{P}_{\max}(\rho_{B})-{\mathcal{C}}(\rho_{B})\leq \mathcal{J}(B|\{\Omega^A _{i}\}),\\
\Delta\mathcal{C}^{T}_{\max}&=&\overline{\mathcal{C}}^{T}_{\max}(\rho_{B})-{\mathcal{C}}^{T}(\rho_{B})\leq \mathcal{J}(B|\{\Omega^A _{i}\}). \label{zongmax}
\end{eqnarray}
with
\begin{equation}
\overline{\mathcal{C}}^{P/T}_{\max}(\rho_{B})=\max_{\Pi}\overline{\mathcal{C}}^{P/T}_{\Pi}(\rho_{B}). \label{definma}
\end{equation}
If $\rho_B$ is incoherent, we have
\begin{equation}
\overline{\mathcal{C}}^{P/T}_{\Pi}(\rho_{B})\leq\overline{\mathcal{C}}^{P/T}_{\max}(\rho_{B})\leq \mathcal{J}(B|\{\Omega^A _{i}\}).\label{zmaxi}
\end{equation}

\textit{Proof.} It is obvious from theorem 1, 2 and 3. \hfill$\blacksquare$

\textbf{Corollary 2}: If $\rho_{AB}$ satisfies   $S(\rho_{B})-S(\rho_{A})=S(\rho_{AB})$, then \begin{equation}
\Delta\mathcal{C}_{\Pi}^{P/T}\leq S(\rho_{A}). \label{30x}
\end{equation}

\emph{Proof.} If the initial quantum state $\rho_{AB}$ satisfies the
  $S(\rho_{B})-S(%
\rho_{A})=S(\rho_{AB})$, we have \cite{Xi85}
\begin{equation}
\mathcal{D}_{A}(\rho_{AB})=S(\rho_{A}).  \label{13x}
\end{equation}
where $\mathcal{D}_{A}(\rho_{AB})$ is the quantum discord defined
  by $\mathcal{D}_{A}(\rho_{AB})=\mathcal{I}(\rho_{AB})-\mathcal{J}(B|\{\Omega^A _{i}\})$ with
$\mathcal{I}(\rho_{AB})=S(\rho_A)+S(\rho_B)-S(\rho_{AB})$. Thus one can easily show $\mathcal{J}(B|\{\Omega^A _{i}\})=S(\rho_A)$ which completes the proof.  \hfill$\blacksquare$

\textbf{Theorem 4.}  Taking all Alice's possible measurements into account, no extra MIAC  is present if and only if the state $\rho_{AB}$ is block-diagonal under Bob's computational basis or a product state.

\emph{Proof.} Consider the computational basis $\{\left\vert i\right\rangle_B\}$, the state $\rho_{AB}=\sum_{ii}M^A_{ii}\otimes\left\vert i\right\rangle_B\left\langle i\right\vert+\sum_{i\neq j}M^A_{ij}\otimes \left\vert i\right\rangle_B\left\langle j\right\vert$ where $M^A_{ii}$ is Hermitian and positive and $\rho_A=\mathrm{Tr}_B\rho_{AB}=\sum_i M^A_{ii}$. It is obvious that if $M^{A}_{ij}=0$ for all $i\neq j$, the states Bob obtains are always diagonal subject to $\{\left\vert i\right\rangle_B\}$. That is, no extra MIAC can be obtained. If $\rho_{AB}$ is a product state which implies $\mathcal{J}(B|\{\Omega^A _{i}\})=0$, it means that the upper bound of the extra MIAC is zero based on Theorem 2. So no extra MIAC could be obtained.

On the contrary, no extra MIAC includes two cases: one is that the final average coherence is zero, and the other is that the final nonzero average coherence is not increased compared with the coherence of $\rho_B=\mathrm{Tr}_A\rho_{AB}$.
The first case means that Alice performs a measurement $\{\left\vert \pi_i\right\rangle_A\left\langle\pi_i\right\vert\}$ (optimal for the maximal average coherence) such that Bob obtains an ensemble $\{p_i,\rho^B_i\}$ where $\rho_B=\sum_i p_i\varrho^B_i$ with  all $\varrho^B_i$  diagonal. Thus  $\rho_{AB}$ can be written as
\begin{equation}
\rho_{AB}=\sum_{i\neq j} p_i\left\vert \pi_i\right\rangle_A\left\langle\pi_i\right\vert\otimes \varrho^B_i+\left\vert \pi_i\right\rangle_A\left\langle\pi_j\right\vert\otimes ({M}^B_{ij}+N^B_{ij}), \label{fanz}
\end{equation}
where ${M}^B_{ij}$ is diagonal and $N^B_{ij}$ has no nonzero diagonal entries. Assume there is at least one nonzero matrix $N^B_{ij}$ among all $i,j$, then one can always select a projector $\left\vert \varphi\right\rangle\left\langle\varphi\right\vert$ such that $\sum_{i\neq j}\left\langle \varphi\right\vert\left.\pi_i\right\rangle\left\langle \pi_j\right\vert\left.\varphi\right\rangle N^B_{ij}\neq 0$.  This means that Bob can get a state with some coherence. In other words,  $\{\left\vert \pi_i\right\rangle_A\left\langle\pi_i\right\vert\}$ is not the optimal measurement, which is a contradiction. So we have $N^B_{ij}=0$. Under this condition, one can find from Eq. (\ref{fanz}) that $\rho_{AB}$ is block-diagonal subject to Bob's basis $\{\left\vert i\right\rangle_B\}$. The second case implies that there exists a decomposition $\{p_i,\rho^B_i\}$ (optimal for the maximal average coherence) with $\rho_B=\sum_i p_i\rho^B_i$ such that $\mathcal{C}(\rho_B)=\sum_i p_i \mathcal{C}(\rho^B_i)$ which, however, is only satisfied when all $\rho^B_i$ are the same for nonzero $\sum_i p_i \mathcal{C}(\rho^B_i)$, since $\mathcal{C}$ is a convex function. Thus we have $\rho^B=\rho^B_i$ which leads to $S(\rho_{B})=\sum_i p_iS(\rho^B_i)$. Now we claim that  $\{p_i,\rho^B_i\}$ is also optimal for the classical correlation. This can be seen as follows. If there exists another decomposition $\{p'_i,\rho'^B_i\}$ for the classical correlation, $\rho'^B_i$ cannot be the same, which will lead to the larger average coherence due to the convexity of $\mathcal{C}$. This is a contradiction. So $\{p_i,\rho^B_i\}$ is the optimal decomposition for the classical correlation, that is, $\mathcal{J}(B|\{\Omega^A _{i}\})=S(\rho_B)-\sum_i p_i S(\rho^B_i)=0$ which implies $\rho_{AB}$ is a product state. The proof is finished. \hfill{}$\blacksquare$

\textbf{Theorem 5.}  Consider all Alice's possible measurements, no extra MIATC is present if and only if the state $\rho_{AB}$ is a product state.

\emph{Proof.} A product state has no classical correlation, i.e., $\mathcal{J}(B|\{\Omega^A _{i}\})=0$ which implies that the upper bound of the extra MIATC is zero in terms of Theorem 3. Thus no extra MIATC could be obtained.

On the contrary, no extra MIATC implies that $\Delta\mathcal{C}^{T}_{\Pi}=\overline{\mathcal{C}}^{T}_{\Pi}(\rho_{B})-\mathcal{C}^{T}(\rho_{B})=S(\rho_{B})-\sum_{i}p_{i}S(\rho_{B}^{i})=0$, namely $S(\rho_{B})=\sum_{i}p_{i}S(\rho_{B}^{i})$. Similar to the proof of theorem 4, one can find that $\mathcal{J}(B|\{\Omega^A _{i}\})=S(\rho_B)-\sum_i p_i S(\rho^B_i)=0$ which corresponds to a product state $\rho_{AB}$. The proof is finished. \hfill{}$\blacksquare$

\textbf{{Examples}-} The above theorems mainly show that, even though the coherence is the quantum feature of a quantum system, in the particular game as sketched in FIG. 1, the extra average coherence obtained by Bob with the assistance of Alice's measurement is well bounded by the classical correlation of their shared state, instead of the quantum correlation. However, one can find that the necessity for all the attainable bounds is to share the  pure states which happen to own the equal quantum and classical correlations. Therefore, one could think that the classical correlation is trivial in contrast to the quantum correlation (e.g., quantum correlation serves as a tight upper bound, but is less than classical correlation ). The following examples show that it is not the case.

\textit{Example.1-The extra average coherence could be induced in classical-classical states.}
Suppose a bipartite state is given by
\begin{eqnarray}
\rho_{AB}=\frac{1}{2}(|0\rangle_{A}\langle 0|\otimes |+\rangle_{B}\langle +|+|1\rangle_{A}\langle 1|\otimes |-\rangle_{B}\langle -|),
\end{eqnarray}
with $|\pm\rangle=\frac{1}{\sqrt{2}}(|0\rangle\pm|1\rangle)$, the reduced quantum state $\rho_{A}=\rho_{B}=\frac{\mathbb{I}_2}{2}$ is incoherent. So the classical correlation is equal to the total correlation, i.e.,
\begin{eqnarray}
\mathcal{J}(\rho_{AB})=S(\rho_{A})+S(\rho_{B})-S(\rho_{AB})=1.
\end{eqnarray}

If the subsystem $A$ is measured by the projective measurements $\Pi:\{|0\rangle\langle 0|,|1\rangle\langle 1|\}$, subsystem B will collapse to the state $\rho_{B}=|\pm\rangle\langle \pm|$ with the probability $p_{\pm}=\frac{1}{2}$. The extra MIAC and the extra MIATC subject to the measurement $\Pi$ can be calculated as
\begin{eqnarray}
\Delta\mathcal{C}_{\Pi}^{P}&=&p_{+}\mathcal{C}_{r}(\rho_{+}^{B})+p_{-}\mathcal{C}_{r}(\rho_{-}^{B})-\mathcal{C}(\rho_{B})=1,\\
\Delta\mathcal{C}_{\Pi}^{T}&=&
\log d-\sum_{i}p_{i}S(\rho_{i}^{B})-\mathcal{C}^{T}(\rho_{B})=1.
\end{eqnarray}
If the subsystem $A$ is measured by the projective measurement $\Pi:\{|+\rangle\langle +|,|-\rangle\langle -|\}$, subsystem B will collapse to the state $\rho_{+}^{B}=\rho_{-}^{B} =\mathbb{I}_2/2$ with the equal probability.
So there is no extra MIAC and MIATC.
This example shows that the extra average coherence is well bounded by the classical correlation. In particular, it also shows that the extra average coherence could exist even though not any quantum correlation is present.

\textit{Example 2. No extra average coherence could be induced in the classical-quantum state.}
Set the classical-quantum state as
\begin{eqnarray}
\rho_{AB}=\frac{1}{2}|0\rangle_{A}\langle 0|\otimes\rho_{1}^{B}+\frac{1}{2}|1\rangle_{A}\langle 1|\otimes\rho_{2}^{B}.
\end{eqnarray}
with $\rho_{1}^{B}=|+\rangle\langle+|$ and $\rho_{2}^{B}=|0\rangle\langle 0|$. The reduced quantum states are given by
\begin{eqnarray}
\rho_{A}=\left(\begin{array}{cc}
          \frac{1}{2} & 0 \\
          0& \frac{1}{2}\\
          \end{array}\right),
\rho_{B}=\left(\begin{array}{cc}
          \frac{3}{4} & \frac{1}{4} \\
          \frac{1}{4}& \frac{1}{4}\\
          \end{array}\right).
\end{eqnarray}
Since there is no quantum correlation subject to subsystem A, the corresponding classical correlation is directly determined by the total correlation as
\begin{eqnarray}
\mathcal{J}(B|\{\Omega _{i}^A\})=S(\rho_{A})+S(\rho_{B})-S(\rho_{AB})=-\sum_{\pm}\frac{2\pm\sqrt{2}}{4}\log \frac{2\pm\sqrt{2}}{4}. \label{jvj}
\end{eqnarray}

Suppose that the projective measurement $\Pi:\{|+\rangle\langle+|,|-\rangle\langle-|\}$ is performed on subsystem A while the subsystem B will collapse on the state
$ \rho_{\pm}^{B}=\left(\begin{array}{cc} \frac{3}{4} & \frac{1}{4} \\ \frac{1}{4}& \frac{1}{4}\\\end{array}\right) $
 with the equal probability $p_{+}=p_{-}=\frac{1}{2}$. It is obvious that there is no extra average coherence ($\Delta\mathcal{C}_{\Pi}^{P/T}=0$) gained by this measurement.
However, if the projective measurement is selected as $\Pi:\{|0\rangle\langle0|,|1\rangle\langle1|\}$, subsystem B will be at the state $\rho_1^B$ and $\rho_2^B$ with the equal probability. Therefore, the \textit{nonzero} extra average coherence can be obtained as
\begin{eqnarray}
\Delta\mathcal{C}_{\Pi}^{P}&=&\mathcal{J}(B|\{\Omega _{i}^A\})+\frac{1}{2}+\frac{1}{4}\log \frac{1}{4}+\frac{3}{4}\log \frac{3}{4},\\
\Delta\mathcal{C}_{\Pi}^{T}&=&\mathcal{J}(B|\{\Omega _{i}^A\}),
\end{eqnarray}
with $\mathcal{J}(B|\{\Omega _{i}^A\})$ given by Eq. (\ref{jvj}). This example shows that  an improper measurement could induce no extra average coherence even though quantum correlation is absent.

\textit{Example 3. No extra average coherence could be induced in the quantum-classical state.}
Suppose the quantum-classical state is given by
\begin{eqnarray}
\rho_{AB}=\frac{1}{2}\rho_{1}^{A}\otimes|+\rangle_{B}\langle+|+\frac{1}{2}\rho_{2}^{A}\otimes|-\rangle_{B}\langle-|,
\end{eqnarray}
with
$
\rho_{1}^{A}=|+\rangle_{A}\langle+|
$ and $\rho_{2}^{A}=\left\vert 0\right\rangle_{A}\left\langle 0\right\vert
$.
It is easy to see that the reduced quantum state $\rho_{B}=\mathbb{I}_2/2$ is incoherent, i.e., $\mathcal{C}(\rho_B)=0$. The classical
correlation is \begin{eqnarray}
\mathcal{J}(B|\{\Omega _{i}^A\})=1+\sum_\pm\frac{2\pm\sqrt{2}}{4}\log \frac{2\pm\sqrt{2}}{4}.\label{jv}
\end{eqnarray}

If the projective measurement $\Pi:\{|0\rangle\langle0|,|1\rangle\langle1|\}$ is used on subsystem A, subsystem B will be on the states $
 \rho_{1}^{B}=\frac{1}{3}\left\vert +\right\rangle\left\langle +\right\vert+\frac{2}{3}\left\vert -\right\rangle\left\langle -\right\vert$ and
 $ \rho_{2}^{B}=\left\vert +\right\rangle\left\langle +\right\vert$ with the corresponding probability  $p_{1}=\frac{3}{4}$ and $p_{2}=\frac{1}{4}$.
Thus a simple calculation can show \begin{eqnarray}
\Delta\mathcal{C}_{\Pi}^{P}=\Delta\mathcal{C}_{\Pi}^{T}=1+\frac{1}{4}\log{\frac{1}{3}}+\frac{1}{2}\log{\frac{2}{3}}.\end{eqnarray}
However, if we select another projective measurement $\Pi:\{\left\vert\psi^\pm(\theta,\phi)\right\rangle\left\langle\psi^\pm(\theta,\phi)\right\vert\}$ where  $|\psi^+(\theta,\phi)\rangle=\cos{\theta}\left\vert 0\right\rangle+e^{i\phi}\sin{\theta}\left\vert 1\right\rangle$ and  $|\psi^-(\theta,\phi)\rangle=\sin{\theta}\left\vert 0\right\rangle-e^{i\phi}\cos{\theta}\left\vert 1\right\rangle$ with $\cot{2\theta}=\cos{\phi}$,  subsystem B will collapse to
 \begin{eqnarray}
 \rho_{\pm}^{B}=a_\pm|+\rangle_{B}\langle+|+b_\pm|-\rangle_{B}\langle-|,\end{eqnarray}
 where $a_\pm=\left\vert\left\langle\psi^\pm(\theta,\phi)\right\vert \left. +\right\rangle\right\vert ^2/p_\pm$ and $b_\pm=\left\vert\left\langle\psi^\pm(\theta,\phi)\right\vert \left. 0\right\rangle\right\vert ^2/p_\pm$ with the probability $p_+=\frac{\left\vert\cos\theta+e^{i\phi}\sin\theta\right\vert^2+\cos^2\theta}{2}$ and $p_-=\frac{\left\vert\sin\theta-e^{i\phi}\cos\theta\right\vert^2+\sin^2\theta}{2}$.
It is easy to demonstrate that ${a_\pm}={b_\pm}$ for  $\cot{2\theta}=\cos{\phi}$ which further leads to  $\rho_{\pm}^{B}=\frac{\mathbb{I}_2}{2}$.
Thus there is no extra average coherence can be gained in terms of this measurement constraint, that is,
\begin{eqnarray}
\Delta\mathcal{C}_{\Pi}^{P}=\Delta\mathcal{C}_{\Pi}^{T}=0.
\end{eqnarray}
Similar to the second example, an improper measurement could induce no extra average coherence even though quantum correlation is present.

\textit{Example 4. The classical correlation can be tighter than the quantum correlation.} Consider a Bell-diagonal state \begin{eqnarray}
\rho_{AB}=\frac{1}{4}\left(\mathbb{I}_{2}\otimes\mathbb{I}_{2}+\sum_{j=1}^{3}c_{j}\sigma_{j}\otimes\sigma_{j}\right),
\end{eqnarray}
where $\vec{\sigma }=(\sigma _{x},\sigma _{y},\sigma _{z})$ is the Pauli matrices. $\rho_{AB}$ is symmetric under exchanging the subsystems. The classical and the quantum correlations are respectively given by  \cite{Luo77}
\begin{eqnarray}
\mathcal{J}(\rho_{AB})&=&\sum_{\pm}\frac{1\pm c}{2}\log(1\pm c),\\
\mathcal{D}_{A/B}(\rho_{AB})&=&\frac{1}{4}[(1-c_{1}-c_{2}-c_{3})\log(1-c_{1}-c_{2}-c_{3}) \nonumber\\
&+&(1-c_{1}+c_{2}+c_{3})\log(1-c_{1}+c_{2}+c_{3}) \nonumber\\
&+&(1+c_{1}-c_{2}+c_{3})\log(1+c_{1}-c_{2}+c_{3}) \nonumber\\
&+&(1+c_{1}+c_{2}-c_{3})\log(1+c_{1}+c_{2}-c_{3})] \nonumber\\
&-&\frac{1-c}{2}\log(1-c)-\frac{1+c}{2}\log(1+c),
\end{eqnarray}
with  $c=\max \{|c_{1}|,|c_{2}|,|c_{3}|\}$.

Suppose the projective measurement $\{\Pi_{\pm}\}:
\Pi_{+}=|\psi(\theta,\varphi)\rangle \langle\psi(\theta,\varphi)|,
\Pi_{-}=1-\Pi_{+}
$
with $|\psi(\theta,\varphi)\rangle=\cos\frac{\theta}{2}|0\rangle+e^{i\varphi}\sin\frac{\theta}{2}|1\rangle$ is performed on subsystem A, subsystem B will collapse, with the equal probability $p_{\pm}=\frac{1}{2}$, on the states \begin{eqnarray}
\rho_{B}^{\pm}&=&\left(
               \begin{array}{cc}
\frac{1\pm c_{3}\bar{s}_\theta }{2}& \pm \frac{s_ \theta}{2}(c_{1}\bar{s}_\varphi-ic_{2} s_\varphi) \\
\pm \frac{s_\theta}{2}(c_{1}\bar{s}_\varphi+ic_{2}s_\varphi) &\frac{ 1\mp c_{3}\bar{s}_\theta}{2} \\
               \end{array}
             \right),\end{eqnarray}
 where $s_x=\sin(x)$ and $\bar{s}_x=\cos(x)$.
 In addition, it is obvious that the reduced quantum states $\rho_{A/B}=\mathbb{I}_2/2$ which implies $\mathcal{C}(\rho_{A/B})=0$.
So the extra average coherence can be directly given by the MIAC or MIATC as
\begin{eqnarray}
&&\Delta{\mathcal{C}}^{P}_{\Pi}(\rho_{B})=\overline{\mathcal{C}}^{P}_{\Pi}(\rho_{B})=\sum_\pm\frac{2\pm\sqrt{\Delta}}{4}\log\frac{2\pm\sqrt{\Delta}}{4}-\frac{1\pm c_{3}\bar{s}_\theta}{2}\log\frac{1\pm c_{3}\bar{s}_\theta}{2}\\
&&\Delta{\mathcal{C}}^{T}_{\Pi}(\rho_{B})=\overline{\mathcal{C}}^{T}_{\Pi}(\rho_{B})=1+\sum_{\pm}\frac{2\pm\sqrt{\Delta}}{4}\log\frac{2\pm\sqrt{\Delta}}{4},
\end{eqnarray}
with $\Delta=c_{1}^{2}+c_{2}^{2}+2c_{3}^{2}-(c_{1}^2+c_{2}^2-2c_{3}^{2})\cos2\theta+2(c_{1}-c_{2})(c_{1}+c_{2})\cos2\varphi\sin^{2}\theta$.

\begin{figure}[t]
\centering
\includegraphics[width=0.5\columnwidth]{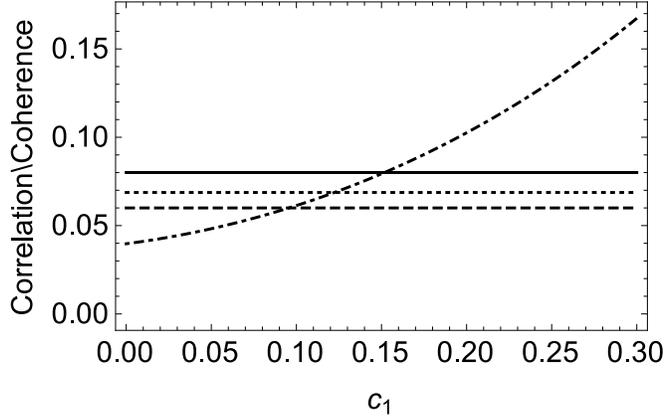}\newline
\caption{(dimensionless) The classical correlation $\mathcal{J}(\rho_{AB})$ (solid line), the quantum correlation $\mathcal{D}_{B}(\rho_{AB})$ (dotted-dashed line), the (extra) MIATC $\overline{\mathcal{C}}^{T}_{\Pi}(\rho_{A})$ (dotted line) and the (extra) MIAC $\overline{\mathcal{C}}^{P}_{\Pi}(\rho_{A})$ (dashed line) versus $c_{1}$ for the Bell-diagonal state.}
\label{fig1}
\end{figure}
In FIG. 2, we plot the quantum and classical correlations and the extra average coherence with the varying $c_1$. The parameters are  chosen as $\theta=2\pi/3,\varphi=\pi/2$ and $c_{2}=0.33,c_{3}=0.22$. The solid line, dotted-dashed line, dotted line and dashed line correspond to the classical correlation, the quantum correlation, the MIATC and the MIAC, respectively. One can find that the classical correlation serves as the good upper bound  for both the (extra) MIATC and the (extra) MIAC and meanwhile, the (extra) MIATC is always greater than the (extra) MIATC. However, the quantum correlation crossing the classical correlation, the (extra)  MIATC and the (extra)  MIAC with the increasing $c_1$ cannot act as a good bound.

\section*{Discussion}

Before the end, we would like to emphasize that all the results in the paper are valid for the POVMs, since it was shown \cite{Datta} that the classical correlations always attained by the rank-one POVM. In addition, we have claimed that Bob isn't allowed to do any operation, which is mainly for the basis-dependent coherence measure. In fact, when we consider the basis-free coherence measure, it is equivalent to allowing Bob to select the optimal unitary operations on his particle. In this case, theorem 3 implies that for pure states the  extra MIATC is  the exact quantum entanglement of their shared state (von Neumann entropy of the reduced density matrix). Thus the coherence also provides an operational meaning for the pure-state entanglement under LOCC.

To sum up, we employ the basis-dependent and basis-free coherence measure to study the extra average coherence induced by a unilateral quantum measurement. Despite that the coherence is the most fundamental quantum feature, we find that the extra average coherence is limited by the classical correlation instead of the quantum correlation. In addition, we find the necessary and sufficient condition for the zero maximal average coherence. We also show that the quantum correlation is neither sufficient nor necessary for the extra average coherence by some examples.

\section*{Methods}

\textbf{{Proof of Theorem 2.}-} We will give the main proof the theorem 2. in the main text. Following Eq. (\ref{MIATCDA}), we have
\begin{eqnarray}
\overline{\mathcal{C}}^{P}_{\Pi}(\rho _{B})-\mathcal{C}(\rho_B)&\leq&S(\rho _{B})-\sum_{i}p_{i}S(\rho ^{B}_{i})\leq \mathcal{J}(B|\{\Omega _{i}^A\}), \label{15}
\end{eqnarray}%
where the second inequality holds due to the optimal $\{\Omega_i^A\}$ implied in Eq. (\ref{jingdian}). So Eq. (\ref{zong}) is satisfied.

In addition, Eq. (\ref{zong}) saturates if both Eq. (\ref{MIATCDA}) and Eq. (15) saturate. Eq. (15) means that $\rho _{i}^{B}$ induced by the measurement $\Pi$ achieves the classical correlation $\mathcal{J}(B|\{\Omega ^A _{i}\})$ and Eq. (\ref{MIATCDA}) implies  $(\rho _{i}^{B})^\star$'s  are the same for all $i$. In order to find an explicit example, suppose $\sigma_{AB}=\left\vert\varphi\right\rangle_{AB}\left\langle\varphi\right\vert$ with
\begin{eqnarray}
|\varphi\rangle_{AB}=\sum_{j}\lambda_{j}U_{A}\otimes\mathbb{I}_B\left\vert j\right\rangle_{A}\left\vert\tilde{j}\right\rangle_{B}, \label{ps}
\end{eqnarray}
with the real $\lambda_{j}$ satisfying $\sum_{j}\lambda_{j}^2=1$. It is obvious
$\sigma_B=\mathrm{Tr}_B\sigma_{AB}=\sum_j\lambda_j^2\left\vert\tilde{j}\right\rangle_B\left\langle\tilde{j}\right\vert$ is incoherent
with respect to the basis $\{|\tilde{j}\rangle\}_{j=1,2,...,d}$. It means
\begin{equation}
\mathcal{C}(\sigma_B)=0.\label{rmb}
\end{equation}
 In order to select a proper measurement, Alice first applies a unitary operation $U'_A$ such that
\begin{eqnarray}
U'_AU_A|j\rangle_{A}=\frac{1}{\sqrt{N}}\sum_{\omega=0}^{N-1}e^{\frac{2\pi ij\omega}{N}}|\omega\rangle_{A},
\end{eqnarray}
with $N$ denoting the dimension of the subsystem A.
Thus $|\varphi\rangle_{AB}$ becomes
\begin{eqnarray}
|\varphi'\rangle_{AB}=\sum_{j}\lambda_{j}\frac{1}{\sqrt{N}}\sum_{\omega=0}^{N-1}e^{\frac{2\pi ij\omega}{N}}|\omega\rangle_{A}|\tilde{j}\rangle_{B}.
\end{eqnarray}
Now Alice performs  the projective measurement $\Omega:\{|\omega\rangle\langle\omega|\}$ on $|\varphi'\rangle_{AB}$, Bob will obtain his state as
\begin{eqnarray}
\sqrt{p_{\omega}}|\phi\rangle_{\omega}^{B}=\sum_{j}\lambda_{j}\frac{1}{\sqrt{N}}e^{\frac{2\pi ij\omega}{N}}|\tilde{j}\rangle_{B},
\end{eqnarray}
with the probability $p_\omega$ corresponding to the measurement outcome $\omega$.
Bob's MIAC can be given by
\begin{eqnarray}
\overline{\mathcal{C}}_{r}(\sigma_{B})&=&\sum_{\omega}p_{\omega}\min_{\delta_{\omega}^{B}}S(\rho_{\omega}^{ B}\|\delta_{\omega}^ {B}) \notag \\
&=&\sum_{\omega}p_{\omega}\left[ S(\rho_{\omega}^{ B^{\star}})-S(\rho_{\omega}^{ B})\right] \notag\\
&=&\sum_{\omega}p_{\omega}(-\sum_{j}\lambda_{j}^{2}\ln \lambda_{j}^{2}) \notag\\
&=&S(\sigma_{B}),\label{phim}
\end{eqnarray}
with $\rho_\omega^B=\left\vert\phi\right\rangle^B_\omega\left\langle\phi\right\vert$.
 For the pure state $\left\vert\varphi\right\rangle_{AB}$, it can prove that the classical correlation $\mathcal{J}(B|\{\Omega _{i}^A\})$ is exactly given by
\begin{equation}
\mathcal{J}(B|\{\Omega _{i}^A\})=S(\sigma _{B})\label{phid}.
\end{equation}%
Eqs. (\ref{rmb}), (\ref{phim}) and (\ref{phid}) show that Eq. (\ref{zong}) saturates for the pure state given by Eq. (\ref{ps}). The proof is finished. \hfill $\blacksquare$

\section*{Acknowledgements (not compulsory)}

This work was supported by the National Natural Science
Foundation of China, under Grant No.11375036, the Xinghai Scholar
Cultivation Plan and the Fundamental Research Funds for the Central
Universities under Grant No. DUT15LK35.

\section*{Author contributions statement}

J.Z. and S.-R.Y. and Y.Z. and C.-S.Y. analyzed the results and wrote the main
manuscript text. All authors reviewed the manuscript.

\section*{Additional information}

Competing financial interests: The authors declare no competing financial
interests.


\begin{thebibliography}{99}

\bibitem{Aberg113} {\AA}berg, J. Catalytic Coherence. \textit{Phys. Rev. Lett.} \textbf{113}, 150402 (2014).

\bibitem{Narasimhachar6} Narasimhachar, V. \& Gour, G. Low-temperature thermodynamics with quantum coherence. \textit{Nat. Commun.} \textbf{6}, 7689 (2015).

\bibitem{Cwiklinski115} \'{C}wikli\'{n}ski, P., Studzi\'{n}ski, M., Horodecki, M. \& Oppenheim, J. Towards fully quantum second laws of thermodynamics: limitations on the evolution of quantum coherences. \textit{Phys. Rev. Lett.} \textbf{115}, 210403 (2015).

\bibitem{Lostaglio6} Lostaglio, M., Jennings, D. \& Rudolph, T. Description of quantum coherence in thermodynamic processes requires constraints beyond free energy. \textit{Nat. Commun.} \textbf{6}, 6383 (2015).

\bibitem{Scully108} Scully, M. O. et al. Quantum heat engine power can be increased by noise-induced coherence. \textit{Proc. Natl. Acad. Sci. U. S. A.} \textbf{108}, 15097 (2011).

\bibitem{Scully299} Scully, M. O., Zubairy, M. S., Agarwal, G. S. \& Walther, H. Extracting Work from a Single Heat Bath via Vanishing Quantum Coherence. \textit{Science} \textbf{299}, 862 (2003).

 \bibitem{levi} Levi, F. \& Mintert, F. A. quantitative theory of coherent delocalization. \textit{New J. Phys.} 16, 033007 (2014).

\bibitem{reb} Rebentrost, P., Mohseni, M. \& Aspuru-Guzik, A. Role of Quantum Coherence and Environmental Fluctuations in Chromophoric Energy Transport. \textit{J. Phys. Chem. B} 113, 9942 (2009).

\bibitem{witt} Witt, B. \& Mintert, F. Stationary quantum coherence and transport in disordered networks. \textit{New J. Phys.} 15, 093020 (2013).

\bibitem{Wang53} Wang, L. \& Yu, C. S. The Roles of a Quantum Channel on a Quantum State. \textit{Int. J. Theor. Phys.} \textbf{53}, 715 (2014).

\bibitem{Plenio10}  Plenio, M. B. \& Huelga, S. F. Dephasing-assisted transport: quantum networks and biomolecules. \textit{New J. Phys.} \textbf{10}, 113019 (2008).

\bibitem{Lloyd302} Lloyd, S. Quantum coherence in biological systems. \textit{J. Phys. Conf. Ser.} \textbf{302}, 012037 (2011).

\bibitem{Huelga54} Huelga, S. F. \& Plenio, M. B. Vibrations, quanta and biology. \textit{Contemp. Phys.} \textbf{54}, 181 (2013).

\bibitem{Baumgratz113} Baumgratz, T., Cramer, M. \& Plenio, M. B. Quantifying coherence. \textit{Phys. Rev. Lett.} \textbf{113}, 140401 (2014).

\bibitem{Girolami113} Girolami, D. Observable measure of quantum coherence in finite dimensional systems. \textit{Phys. Rev. Lett.} \textbf{113}, 170401 (2014).

\bibitem{Pires91} Pires, D. P., C\'{e}leri, L. C. \& Soares-Pinto, D. O. Geometric lower bound for a quantum coherence measure. \textit{Phys. Rev. A} \textbf{91}, 042330 (2015).

\bibitem{Shao91} Shao, L. H., Xi, Z. J., Fan, H. \& Li, Y. M. Fidelity and trace-norm distances for quantifying coherence. \textit{Phys. Rev. A} \textbf{91}, 042120 (2015).

\bibitem{Rana93} Rana, S., Parashar, P. \& Lewenstein, M. Trace-distance measure of coherence. \textit{Phys. Rev. A} \textbf{93}, 012110 (2016).

\bibitem{Zhang93} Zhang, Y. R., Shao, L. H., Li, Y. M. \& Fan,H. Quantifying coherence in infinite-dimensional systems. \textit{Phys. Rev. A} \textbf{93}, 012334 (2016).

\bibitem{Winter116} Winter, A. \& Yang,D. Operational resoures theory of coherence. \textit{Phys. Rev. Lett.} \textbf{116}, 120404 (2016).

\bibitem{Yu80} Yu, C. S. \& Song, H. S. Bipartite concurrence and localized coherence. \textit{Phys. Rev. A} \textbf{80}, 022324 (2009).

\bibitem{Hu} Hu, X. Y. \& Fan, H. Extracting quantum coherence via steering. \textit{Sci. Rep.} \textbf{6}, 34380 (2016).

\bibitem{Chitambar116} Chitambar, E. et al. Assisted distillation of quantum coherence. \textit{Phys. Rev. Lett.} \textbf{116}, 070402 (2016).

\bibitem{tan} Tan, K. C., Kwon, H., Park, C.-Y. \& Jeong, H. Unified view of quantum correlations and quantum coherence. \textit{Phys. Rev. A} \textbf{94}, 022329 (2016).

\bibitem{Yu13} Yu, C. S., Zhang, Y. \& Zhao, H. Q. Quantum correlation via quantum coherence. \textit{Quant. Inf. Proc.} \textbf{13}, 1437 (2014).

\bibitem{Yao92} Yao, Y., Xiao, X., Ge, L. \& Sun, C. P. Quantum coherence in multipartite systems. \textit{Phys. Rev. A} \textbf{92}, 022112 (2015).

\bibitem{Xi5} Xi, Z. J., Li, Y. M. \& Fan, H. Quantum coherence and correlation in quantum system. \textit{Sci. Rep.} \textbf{5}, 10922 (2015).

\bibitem{Cheng92} Cheng, S. M. \& Hall, M. J. W. Complementarity relations for quantum coherence. \textit{Phys. Rev. A} \textbf{92}, 042101 (2015).

\bibitem{Singh91} Singh, U., Bera, M. N., Dhar, H. S. \& Pati, A. K. Maximally coherent mixed states: Complementarity between maximal coherence and mixedness. \textit{Phys. Rev. A} \textbf{91}, 052115 (2015).

\bibitem{Singh93} Singh, U., Zhang, L. \& Pati, A. K. Average coherence and its typicality for random pure state. \textit{Phys. Rev. A} \textbf{93}, 032125 (2016).

\bibitem{Du91} Du, S. P., Bai, Z. F. \& Guo, Y. Conditions for coherence transformations under incohernet operations. \textit{Phys. Rev. A} \textbf{91}, 052120 (2015).

\bibitem{Streltsov115} Streltsov, A. et al. Measuring quantum coherence with entanglement. \textit{Phys. Rev. Lett.} \textbf{115}, 020403 (2015).

\bibitem{Ma116} Ma, J. J. et al. Coverting coherence to quantum correlation. \textit{Phys. Rev. Lett.} \textbf{116}, 160407 (2016).

\bibitem{Yuquantum} Yu, C. S., Yang S. R. \& Guo, B. Q. Total quantum coherence and its applications. \textit{Quantm Inf. Process.} \textbf{15}, 3773 (2016).

\bibitem{Henderson34} Henderson, L. \& Vedral, V. Classical, quantum and total correlations. \textit{J. Phys. A: Math. Gen.} \textbf{34}, 6899 (2001).

\bibitem{Xi85} Xi, Z. J., Lu, X. M., Wang, X. G. \& Li, Y. M. Necessary and sufficient condition for saturating the upper bound of quantum discord. \textit{Phys. Rev. A} \textbf{85}, 032109 (2012).

\bibitem{Luo77} Luo, S. L. Quantum discord for two-qubit systems. \textit{Phys. Rev. A} \textbf{77}, 042303 (2008).

\bibitem{Datta} Datta, A. Studies on the Role of Entanglement in Mixed-state Quantum Computation. \textit{arXiv:} 0807.4490 [quant-ph].

\end{thebibliography}
\end{document}